\documentclass[useAMS,usenatbib]{mnras}

\usepackage{mathptmx}
\usepackage{ amssymb }
\usepackage{multirow}
\usepackage{psfrag}
\usepackage{siunitx}
\usepackage{pspicture}
\usepackage{graphicx}
\usepackage{amsmath}
\usepackage{color}
\usepackage{url}
\usepackage{threeparttable}
\usepackage{caption}
\usepackage{lscape}
\usepackage{booktabs}
\usepackage{subcaption}
\usepackage{grffile}
\usepackage{multirow}
\usepackage{multicol}
\usepackage{dcolumn}
\usepackage{hyperref}
 
\usepackage{setspace}
\usepackage{xspace}

\newcommand{\C}{C{\sc iii}]\xspace} 
\newcommand{\CC}{C{\sc ii}]\xspace}
\newcommand{\CIV}{C{\sc iv}\xspace}
\pubyear{2017}

\title[{C{\sc ii}], C{\sc iii}] and C{\sc iv} emitters at $z\sim0.7-1.5$}]{A ${\bf 1.4}$ deg${\bf ^2}$ blind survey for C{\sc ii}], C{\sc iii}] and C{\sc iv} at ${\bf z\sim0.7-1.5}$. II: \\
luminosity functions and cosmic average line ratios}
\author[A. Stroe et al.]{Andra Stroe$^{1}$\thanks{E-mail: astroe@eso.org}\thanks{ESO Fellow}, David Sobral$^{2,3}$, Jorryt Matthee$^{2}$, Jo\~ao Calhau$^{3}$, Ivan Oteo$^{4,1}$\\
$^{1}$European Southern Observatory, Karl-Schwarzschild-Str. 2, 85748, Garching, Germany\\
$^{2}$Leiden Observatory, Leiden University, P.O.\ Box 9513, NL-2300 RA Leiden, The Netherlands\\
$^{3}$Department of Physics, Lancaster University, Lancaster, LA1 4YB, UK\\
$^{4}$Institute for Astronomy, University of Edinburgh, Royal Observatory, Blackford Hill, Edinburgh EH9 3HJ UK}
\begin{document}

\maketitle
\begin{abstract}
Recently, the \C and \CIV emission lines have been observed in galaxies in the early Universe ($z>5$), providing new ways to measure their redshift and study their stellar populations and AGN. We explore the first blind \CC, \C and \CIV survey ($z\sim0.68, 1.05, 1.53$, respectively) presented in Stroe et al. (2017). We derive luminosity functions (LF) and study properties of \CC, \C and \CIV line emitters through comparisons to the LFs of H$\alpha$ and Ly$\alpha$ emitters, UV selected star forming (SF) galaxies and quasars at similar redshifts. The \CC LF at $z\sim0.68$ is equally well described by a Schechter or a power-law LF, characteristic of a mixture of SF and AGN activity. The \C LF ($z\sim1.05$) is consistent to a scaled down version of the Schechter H$\alpha$ and Ly$\alpha$ LF at their redshift, indicating a SF origin. In stark contrast, the \CIV LF at $z\sim1.53$ is well fit by a power-law, quasar-like LF. We find that the brightest UV sources ($M_{UV}<-22$) will universally have \C and \CIV emission. However, on average, \C and \CIV are not as abundant as H$\alpha$ or Ly$\alpha$ emitters at the same redshift, with cosmic average ratios of $\sim0.02-0.06$ to H$\alpha$ and $\sim0.01-0.1$ to intrinsic Ly$\alpha$. We predict that the \C and \CIV lines can only be truly competitive in confirming high redshift candidates when the hosts are intrinsically bright and the effective Ly$\alpha$ escape fraction is below 1 per cent. While \C and \CIV were proposed as good tracers of young, relatively low-metallicity galaxies typical of the early Universe, we find that, at least at $z\sim1.5$, \CIV is exclusively hosted by AGN/quasars, especially at large line equivalent widths. 
\end{abstract}

\section{Introduction}\label{sec:intro}

The star formation (SF) rate density of the Universe grows significantly from $z\sim0$, reaching a peak at $z\sim2-3$, but is then measured to decline steeply into the epoch of reionisation \citep[e.g.][]{1996ApJ...460L...1L, 1996MNRAS.283.1388M, 2006ApJ...651..142H,2011ApJ...737...90B,2015MNRAS.452.3948K}. Quasars undergo a similar number density evolution: the density of quasars increases up to $z\sim1-3$, only to plummet at higher redshifts \citep[e.g.][]{1990MNRAS.247...19D, 1994ApJ...421..412W, 2006AJ....131.2766R, 2013ApJ...768..105M}.

\renewcommand{\arraystretch}{1.2}
\begin{table*}
\begin{center}
\caption{The three emission lines we study in the present work. We list the restframe wavelength, the ionisation energy \citep[$\chi$,][]{2002ASPC..284..111V}, the redshift range over which the emitters are selected, the average luminosity distance over the redshift range ($D_\mathrm{L}$) and volume at each redshift slice. The final number of sources of each emitter type includes the secure sources with $z_{\rm phot}$ and $z_{\rm spec}$ selected and described in \citet{PaperI}, as well as sources added with fractions (see Section \ref{sec:fractions}). The physical origin of \CC, \C and \CIV emission based on literature and \citet{PaperI} is shown in the last column.}
\begin{tabular}{l c c c c c c c c p{4cm}}
\hline\hline
Line & $\lambda_\mathrm{line}$ & $\chi$ & $z_\mathrm{line}$ & $D_\mathrm{L}$ &  Volume & $z_{\rm spec}$ & $z_{\rm phot}$ & All & Comments\\ 
              & (\AA)                   & (eV) & at FWHM           & ($10^3$ Mpc)          & ($10^5$\,Mpc$^3$) & &  (without $z_{\rm spec}$) &  \\ \hline
\CC & 2326 & 11.3 & $0.673-0.696$ & \phantom{0}4.14 & $1.76$ & \phantom{0}3 & 13 & 22 & SF at lower luminosities, AGN at higher luminosities\\
\C & 1907, 1909 & 24.4 & $1.039-1.066$ & \phantom{0}7.04 & $3.36$  & \phantom{0}4 & 30 & 43 & mostly produced in SF galaxies \\ 
\CIV & 1549, 1551 & 47.9 &   $1.513-1.546$ & 11.17 & $5.29$        & 14 & \phantom{0}3 & 28 & almost exclusively trace quasars \\
\hline
\end{tabular}
\label{tab:lines}
\end{center}
\end{table*}
\renewcommand{\arraystretch}{1.1}

To track the evolution of galaxies across cosmic time, ideally one would use a single tracer of SF activity. Intrinsically the brightest emission line in H{\sc ii} regions, the Ly$\alpha$ line has been traditionally associated to SF activity and has a high excitation \citep[13.6 eV,][]{2002ASPC..284..111V}. However, Ly$\alpha$ is scattered by neutral hydrogen, making it easily absorbed by dust and difficult to escape the host galaxy \citep[for a review see][]{2015PASA...32...27H,2016MNRAS.458..449M, Sobral2017}. While Ly$\alpha$ emitters are typically thought to be low mass, blue, star-forming galaxies, AGN can also be powerful Ly$\alpha$ sources. Indeed, there is mounting evidence that a large fraction of luminous Ly$\alpha$ emitters are powered by AGN, especially at $z<3$ \citep[][]{2008ApJS..176..301O, 2009A&A...498...13N, 2010ApJ...711..928C,Matthee2017}. 

Across redshifts, Ly$\alpha$ line has been widely used to select both SF galaxies and AGN. Ly$\alpha$ has also been the prime way to spectroscopically confirm high-redshift candidates \citep[e.g.][]{2012ApJ...744...83O, 2015ApJ...808..139S, 2015ApJ...810L..12Z} and is used to obtain large samples through the narrow band (NB) technique \citep[e.g.][]{2008ApJS..176..301O, 2014ApJ...797...16K, 2015MNRAS.451..400M, 2015ApJ...809...89T, 2016ApJ...823...20K, 2016MNRAS.463.1678S}. However, only a fraction of the emitter population selected through the NB technique are actual Ly$\alpha$ at high redshift, while the remaining line emitters can be low-z contaminants. Historically, [O{\sc iii}] and [O{\sc ii}] were considered the most important contaminants for Ly$\alpha$ surveys at $z>3$ \citep[e.g.][]{2008ApJS..176..301O, 2014MNRAS.440.2375M}, because they can have high (observed) equivalent widths (EW). However, this is much less of an issue for Ly$\alpha$ surveys at $z\sim2-3$, because the volume for e.g. [O{\sc ii}] is very small. The most notable contaminants for $z\sim2-3$ Ly$\alpha$ searches instead are C{\sc ii}]$_{2326}$ (from now on \CC), C{\sc iii}]$_{1907,1909}$ (from now on \C) and C{\sc iv}$_{1549,1551}$ (from now on \CIV) emitters \citep[][]{Sobral2017}. Additionally, interpreting Ly$\alpha$-selected samples and studying their properties can be challenging because of Ly$\alpha$ resonant scattering. It is therefore difficult to obtain physical characteristics of galaxies from Ly$\alpha$, such as correlating Ly$\alpha$ luminosity with a SF rate (SFR) or BH accretion rate (BHAR) \citep[e.g.][ Calhau et al. in prep.]{2016MNRAS.458..449M}. 

The Ly$\alpha$ escape fraction at fixed radius drops sharply towards the highest redshifts \citep[over the $z\sim6-7$ range, e.g.][]{2016ApJ...827L..14T}, which is explained by scattering through a partially neutral intergalactic medium \citep[e.g.][]{2013ApJ...775L..29T, 2014PASA...31...40D}, making Ly$\alpha$ emission much more extended \citep{2016MNRAS.463.1678S}. The scattering of Ly$\alpha$ at the very highest redshifts effectively means Ly$\alpha$ slit spectroscopy might not be the best choice for confirming $z>6$ candidates selected with the Lyman-break technique. For example, \citet{2014A&A...569A..78V} invested $>50$ h of Very Large Telescope time on a single `normal' Lyman-break galaxy without any detection of Ly$\alpha$. Most high redshift candidates are selected from deep, but small area fields, and are not bright enough in emission lines to be followed up efficiently with spectroscopy. 

After being detected in a handful of high redshift sources (up to $z\sim6-8$, all of which were also Ly$\alpha$ emitters), \C \citep[ionisation potential of 24.4 eV, ][]{2002ASPC..284..111V} and \CIV (47.9 eV) were proposed by \citet{2015MNRAS.450.1846S} as an alternative way to identify galaxies at the highest redshifts ($z>6$) with upcoming telescopes such as the \textit{James Webb Space Telescope}. Therefore, even though \C and \CIV lines are on average weaker than Ly$\alpha$, they seem to be sufficiently prominent in young, sub-solar metallicity ($0.3Z_\odot$) galaxies expected at high redshift and will not suffer from scattering by neutral hydrogen at $z>6$, thus boosting the observed \C/Ly$\alpha$ and \CIV/Ly$\alpha$ ratios (for single-star, `normal' stellar populations at solar metallicity). 

Theory predicts that carbon emission lines, such as \CC, \C and \CIV, should mainly be produced in the broad line region of active galactic nuclei \citep[AGN, ][]{2006agna.book.....O}. However more recent work suggests a different origin for these lines. Models and observations show that \C and \CIV are the brightest UV lines after Ly$\alpha$ in SF galaxies at redshift $z\gtrsim1$ \citep{2003ApJ...588...65S, 2014MNRAS.445.3200S, 2016MNRAS.462.1757G,2016MNRAS.456.3354F}. While \C is mainly fostered in lower-metallicity, lower-mass SF galaxies or starbursts \citep{2014ApJ...790..144B, 2015ApJ...814L...6R, 2016ApJ...833..136J,2017ApJ...838...63D}, \CIV can in principle be produced by massive stars in a very young SF galaxy \citep{2014MNRAS.445.3200S, 2017ApJ...836L..14M, 2017ApJ...839...17S}. 

However, our knowledge of the statistical properties of \CC, \C and \CIV emitters is still very limited since observations mostly targeted either lensed sources, spectroscopically selected sources or sources whose redshift was already known from Ly$\alpha$. Furthermore, no sources have been found at high redshift using just the \C or \CIV line emission. If the \C and \CIV lines are to be used in the future to select high redshift galaxies, we should also aim to understand what they actually trace and how strong we can expect them to be. It is thus crucial to unveil their luminosity functions (LF) and cosmic evolution of these emitters. 

We have embarked on a project to survey \CC, \C and \CIV emitters in a blind, uniform way, over the COSMOS and UDS field. In \citet{PaperI}, we study the properties of individual \CC, \C and \CIV sources and characterise their nature. We find that \CC emission at $z\sim0.68$ is produced in disky, SF galaxies at fainter fluxes, while at larger fluxes \CC is triggered in Seyfert-like galaxies, with a stellar disk and AGN core. Our work unveils that \C emitters have SF morphologies and have UV and optical colours consistent with a general SF population, while \CIV emitters are all young, blue, actively-accreting quasars.

After presenting our sample in \citet{PaperI} and discussing its reliability and source properties, in this paper (Paper II), we focus on the other statistical properties of \CC, \C and \CIV emitters. We present the first LFs, which enable us to compare the number densities of \CC, \C and \CIV emitters with the distribution of H$\alpha$ and Ly$\alpha$ emitters, UV-selected galaxies and quasars. We also discuss the implications of these high ionisation lines at high redshift and provide cosmic average ratios.

Our paper is structured in the following way: in Sections \ref{sec:data}, \ref{sec:select} and \ref{sec:methods} we present the sample of \CC, \C and \CIV emitters, while in Section~\ref{sec:LF} we study the LFs and compare them to the H$\alpha$, Ly$\alpha$, galaxy and quasar LFs. In Section~\ref{sec:discussion} we discuss the implications of our statistical results for \CC, \C and \CIV detections. Our conclusions can be found in Section~\ref{sec:conclusion}.

We use a flat $\Lambda$CDM cosmology with $H_{0}=70$\,km\,s$^{-1}$\,Mpc$^{-1}$, $\Omega_M=0.3$ and $\Omega_{\Lambda}=0.7$. Magnitudes are in the AB system.

\section{Parent sample}\label{sec:data}

The sample of \CC, \C and \CIV emitters is drawn from the CALYMHA survey, which was designed to mainly study Ly$\alpha$ emitters at $z\sim2.23$ \citep{2016MNRAS.458..449M, Sobral2017}. The data we are using comes from a NB filter (NB392, central wavelength $\lambda_C = 3918$\,{\AA} and width $\Delta \lambda = 52$\,{\AA}) mounted on Wide Field Camera on the Isaac Newton Telescope\footnote{\url{http://www.ing.iac.es/Astronomy/telescopes/int/}} to survey an area of $\sim1.4$ deg$^2$ across the COSMOS and UDS fields. CALYMHA captured many lines at lower redshift apart from the target Ly$\alpha$, down to a limiting $3\sigma$ flux of $\sim4\times10^{-17}$ erg\,s$^{-1}$ cm$^{-2}$ and a limiting observed EW limit of 16\,{\AA}. As was also noted in \citet{Sobral2017}, in \citet{PaperI}, we demonstrated that a good fraction of the emitters selected through the NB survey are actually \CC, \C and \CIV emitters. Apart from Ly$\alpha$, \CC, \C and \CIV, the sample also contains emitters such as [O{\sc ii}] ($z\sim0.05$), Mg{\sc i} and Mg{\sc ii}] ($z\sim0.4$) in lower numbers \citep[for details see][]{Sobral2017}.

\begin{figure}
\centering
\includegraphics[trim=0cm 0cm 0cm 0cm, width=0.479\textwidth]{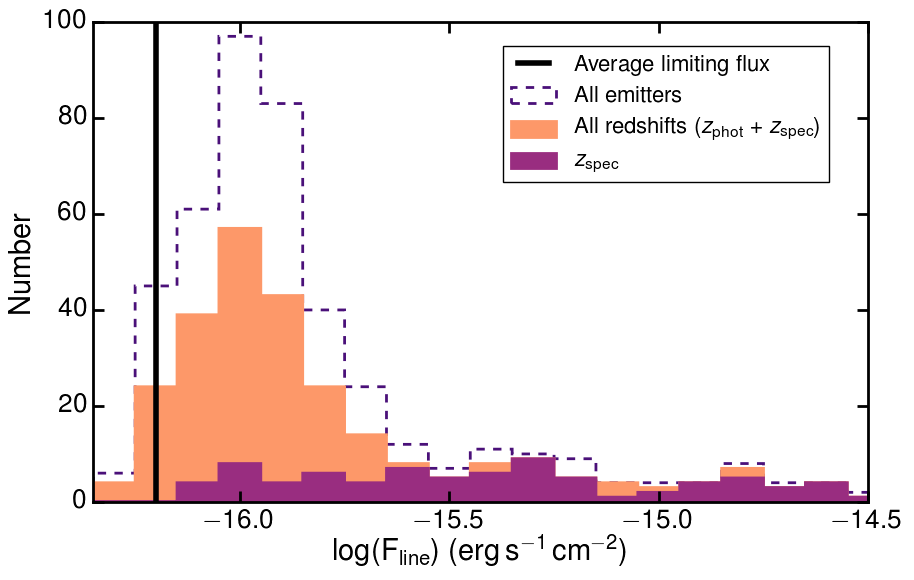}
\caption{Distribution of emitters with respect to line flux. Overplotted is the distribution of all emitters with photometric redshift and those with spectroscopy. Note that we are almost spectroscopically complete at bright fluxes. At lower fluxes, a high fraction of sources have either spectroscopic or photometric redshifts.}
\label{fig:hist}
\end{figure}

\section{Selecting \CC, \C and \CIV emitters}\label{sec:select}

In \citet{PaperI}, we selected \CC, \C and \CIV emitters using spectroscopic and photometric redshifts \citep[see Table~\ref{tab:lines};][]{2009ApJ...690.1236I, 2010MNRAS.401.1166C}. It is important to note that our selection was favouring purity of the sample, rather than completeness. We therefore had conservative photometric redshift ranges and removed any sources that were chosen through colour-colour selections as Ly$\alpha$ by \citet{Sobral2017}. 

Our final sample includes $16$ \CC, $34$ \C and $17$ \CIV emitters chosen based on $z_{\rm spec}$ or $z_{\rm phot}$. The nature of sources without spectroscopic or photometric redshifts is uncertain (a total of 171 sources). Photometric redshifts are not available, for example, for sources which are faint in the continuum, sources in masked areas around bright stars in deep optical data (in the ancillary catalogues) or for faint sources with unusual colours because they are AGN. Deriving photometric redshifts for AGN is challenging. In the COSMOS field, the availability of medium band filters increased the reliability of photometric redshift for AGN powered sources such as some of the \CC emitters and \CIV. However, in UDS such medium band filter measurements are not available to help constrain the $z_{\rm phot}$ fit. 

In order to increase our completeness for the purpose of building reliable LFs, in addition to our secure sources with $z_{\rm phot}$ and $z_{\rm spec}$, we need to estimate how many of sources without colour information might be \CC, \C or \CIV emitters. We therefore derive fractions to describe the probability of a source to be a \CC, \C or \CIV when no secure redshift information is available. In building emitter LFs, these fractions are employed when adding sources to luminosity bins (see Section~\ref{sec:methods}), but we also show our results are not sensitive (within error bars), when we restrict the analysis to the most secures sources. 

\begin{figure}
\centering
\includegraphics[trim=0cm 0cm 0cm 0cm, width=0.479\textwidth]{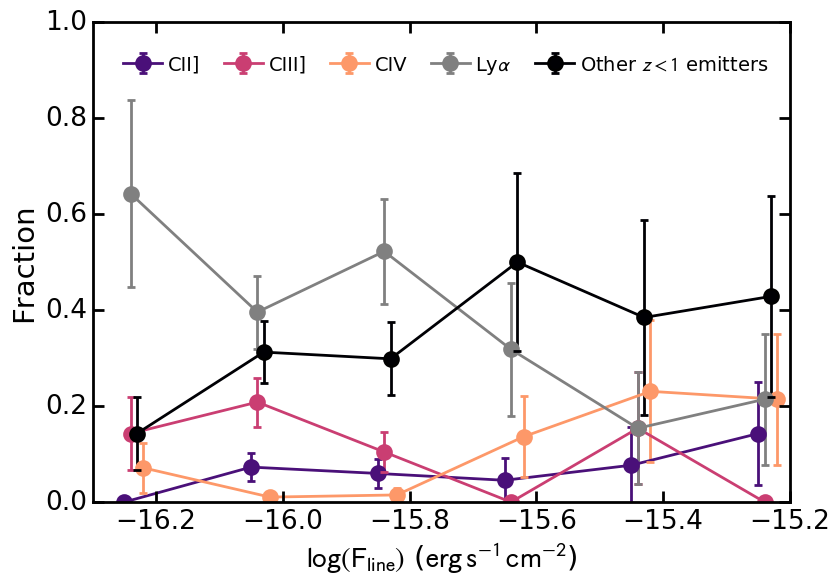}
\caption{The fraction of \CC, \C, \CIV and Ly$\alpha$ emitters as function of line flux, derived using sources with redshifts, down to an observed EW limit of $16$\,{\AA}. The fractions are calculated by dividing the number of emitters of a particular type to the total number of line emitters. Note the decreasing fraction of Ly$\alpha$ emitters at the largest fluxes, where the population is comprised of $\sim40$ per cent \C and \CIV emitters. We also overplot the fraction of lower redshift ($z<1$) emitters such as [O{\sc ii}], Ne{\sc v}, Mg{\sc i} and Mg{\sc ii}]. Note that some of these might still be \CC or \C emitters, given our conservative photometric redshift ranges \citep[see for example, Fig. 2 in][]{PaperI}.}
\label{fig:frac}
\end{figure}

\subsection{\CC, \C, \CIV fractions as function of observed flux}\label{sec:fractions}

By using the photometric and spectroscopic redshifts, we study how the fraction of emitters which are \CC, \C and \CIV depends on the observed line flux, in a similar fashion to the statistical method of \citet{2014MNRAS.438.1377S}. The fraction of a particular C emitter is calculated as the number of this particular emitter type in a flux bin divided by the total number of emitters including all lines. For example for \CC emitters in a particular bin, this is calculated as $N_{\rm CII]}/(N_{\rm CII]}+N_{\rm CIII]}+N_{\rm CIV}+N_{\rm Ly\alpha}+N_{\rm others})$. At the brightest fluxes we are close to spectroscopic completeness ($F_\mathrm{line}\gtrsim10^{15.5}$ erg\,s$^{-1}$\,cm$^{-2}$), while at the fainter fluxes a large fraction of sources has photometric or spectroscopic redshifts (see Fig.~\ref{fig:hist}). This investigation is most relevant for faint sources, to attain a higher completeness in LFs at faint fluxes and avoid selection incompleteness.

We separate the emitter population into secure \CC, \C and \CIV and group the other emitters together. We note that some of the `other' emitters could still be C emitters, given our conservative redshift cuts. We focus on the fluxes below $\sim10^{-15}$ erg\,s$^{-1}$\,cm$^{-2}$, where the number statistics in each flux bin are good enough to derive reliable fractions (Fig.~\ref{fig:frac}). At large fluxes of $>10^{-15.5}$ erg\,s$^{-1}$\,cm$^{-2}$, Carbon species represent $40-50$ per cent of the population \citep[see also][]{Sobral2017}. There is a trend of increasing \CC and \CIV fractions and almost constant \C fraction with increasing line flux. We extrapolate the fractions for the few sources without redshifts at fluxes $>10^{-15.2}$ erg\,s$^{-1}$\,cm$^{-2}$. 

Fractions are only valid for studying statistical, average properties such as LFs, and cannot be used to describe properties of individual sources. Our results depend little on the way we extrapolate the results to higher luminosities or if we vary the fractions within the error bars. We tested a few ways of extrapolating at bright luminosities: we applied a function fit to the available bins, we used the values in the bins directly and for the bright luminosities simply used the value in the brightest bin and also applied fractions to all the UDS emitters, given their less reliable photometric redshifts in classifying the emitters (see also Section~\ref{sec:select}). We found that this did not significantly affect the results and use the function fit to the fractions when deriving LFs.

\section{Methods}\label{sec:methods}

We derive the line luminosity for the \CC, \C and \CIV sources from the line flux:
\begin{equation}
\label{eq:L}
L_\mathrm{line} = 4 \pi D^2_{L}(line) F_\mathrm{line},
\end{equation}
where line is \CC, \C or \CIV and $D_{L}(line)$ is the luminosity distance at the redshift of each line of interest (see Table~\ref{tab:lines}). We note that some of our emitters have large EWs possibly caused by offsets between the main line emitting regions and the underlying stellar light \citep{PaperI}. However, one of the main aims of this work is to study LFs and our samples are ideal for this purpose, since they are flux/luminosity selected, irrespective of the continuum. Therefore, the measurement of fluxes and luminosities is a more reliable measurement compared to EW, as it does not depend on geometrical effects in the same way the EW estimates do \citep[see discussion in][]{PaperI}.

For the purpose of building LFs, we bin the emitters based on their luminosity. For the sources with either photometric or spectroscopic redshifts, we add them to the bin with a weight of 1. For the other emitters, the sources were added with a weight according to the fractions from Fig.~\ref{fig:frac}. The total number of line emitters, obtained by adding the number of secure \CC, \C and \CIV emitters as well as the number obtained through the fractions, is listed in Table~\ref{tab:lines}.
\begin{figure}
\centering
\includegraphics[trim=0cm 0cm 0cm 0cm, width=0.479\textwidth]{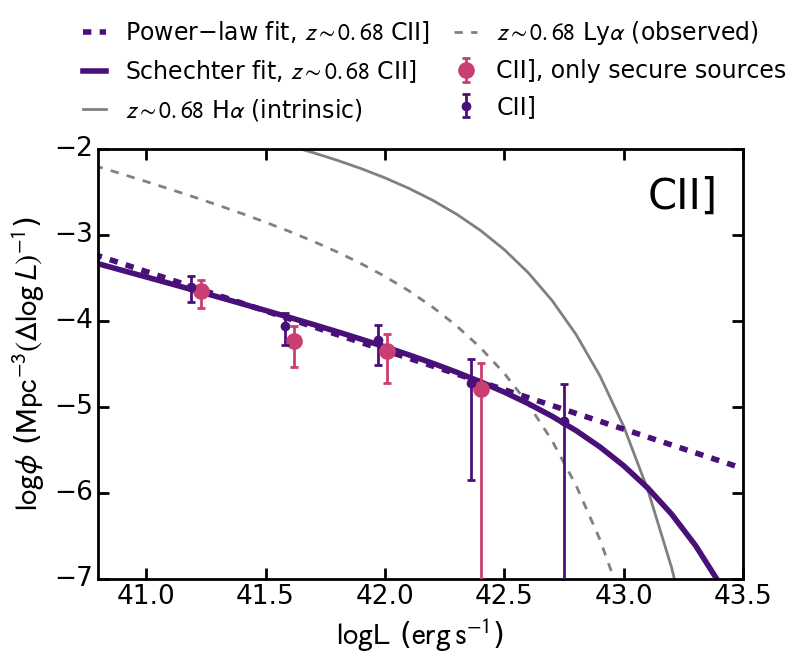}
\caption{The \CC LF at $z\sim0.68$. We also plot the bins obtained with secure sources, which does not result in significantly different fits. The bins are shifted to the right for clarity. Overplotted is the observed Schechter Ly$\alpha$ LF at $z\sim0.68$, interpolated from data at $z\sim0.3-3.1$ \citep{2008ApJS..176..301O, 2010ApJ...711..928C,2012ApJ...749..106B, 2016ApJ...823...20K, Sobral2017}. We also plot the H$\alpha$ LF interpolated from results of \citet{2015MNRAS.453..242S} and \citet{2013MNRAS.428.1128S} to the redshift we trace for \CC in our survey. Our \CC data can be fit with either a Schechter function or a power-law. The LFs indicate that at most luminosities below $L^*$, \CC number densities are a factor of $\sim10$ lower than Ly$\alpha$ and a factor of $\sim100$ lower than H$\alpha$.}
\label{fig:LFCII}
\end{figure}

We derive LFs for the \CC, \C and \CIV emitters down to 30 per cent completeness. We correct the LF for incompleteness using the completeness curves as function of line flux from \citet{Sobral2017}, as well as correct the volumes for the real shape of the filter profile, using the method described in \citet{2012MNRAS.420.1926S} and \citet{Sobral2017}. 

We obtain number density values by dividing the binned numbers of sources by the correct cosmic volume at the emitter redshift (see Table~\ref{tab:lines}). To test the effects of the binning choice, we also resample the data with random choices of bin centres and widths. We settle on the binning choice which best reproduces the average shape of the individual binning choices. The final choice of bin width for all three emitter types is $\log L=0.4$, given the low number statistics.

We use a \citet{1976ApJ...203..297S} function to fit the number densities using a least-squares approach. The errors are Poissonian, with 20 per cent added to account for imperfect fractions and for completeness and filter profile correction errors.
\begin{equation}
\label{eq:schechter}
\phi(L) \mathrm{d} L = \phi^*\left(\frac{L}{L^*}\right)^{\alpha} e^{-\frac{L}{L^*}} \mathrm{d} \left(\frac{L}{L^*}\right),
\end{equation}
where $\alpha$ is the faint end slope of the LF, $\phi^*$ is the characteristic number density of the emitters and $L^*$ is the characteristic emitter luminosity. Given the depth of our data, we also fix the faint end slope of the fit $\alpha$ to $-1.75$. A steep faint end slope of $\sim-1.7$ was found appropriate for the Ly$\alpha$ LF though-out redshifts since $z\sim3$ as well as the H$\alpha$ LF at $z\gtrsim0.7$ \citep[e.g.][]{2010ApJ...711..928C, 2013MNRAS.428.1128S,2016A&A...591A.151G, 2016ApJ...823...20K, Sobral2017}. 

We also fit with a power-law, when the data enables it:
\begin{equation}
\label{eq:powerlaw}
\log \phi(L) = \gamma \log L + \log L_{\rm int},
\end{equation}
where $\gamma$ is the slope of the power-law and $L_{\rm int}$ is the intercept at 0 number density.

Note that a power-law fit to the fainter luminosity bins should be consistent with the faint-end slope of the Schechter fit. Because of the different definitions of the two functions the relationship between the two slopes is $\alpha= \gamma-1$.

As mentioned before (Section~\ref{sec:fractions}), we tested a few ways to set the weights for the sources with fractions (direct values, interpolation, fitting a curve) and found no significant influence on the LF: the fit parameters for all the choices were within error bars. We also produced LFs using only the COSMOS data (which has a volume about 7 times than UDS) and found that removing the UDS data does not affect the results. 

Table~\ref{tab:LF} contains the best-fit Schechter LF parameters and Table~\ref{tab:LFpowerlaw} the power-law fit parameters for the \CC, \C and \CIV emitters.

\section{Luminosity functions}\label{sec:LF}

Here we present the \CC, \C and \CIV LFs and compare theme with the H$\alpha$, Ly$\alpha$, galaxy and quasar UV LFs. We use such comparisons to further investigate the nature of the sources as a whole.

\subsection{\CC luminosity function}\label{sec:LFCII}

We fit our \CC LF with a Schechter function (reduced $\chi^2_{\rm red}=0.3$, see Fig.~\ref{fig:LFCII}), however we cannot constrain the bright end very well, as the usual Schechter number density drop might be located beyond luminosities we can directly probe ($L_{\rm CII]}>10^{42-43}$ erg\,s$^{-1}$). Note the low number statistics in these bright regimes: we have an equivalent of $\sim3-4$ sources between $10^{42}$ and $10^{42.7}$ erg\,s$^{-1}$. Sources added through fractions cannot be responsible for the high densities at these bright luminosities, as 2 sources are spectroscopically confirmed and 1 has a photometric redshift. The bright end behaviour therefore does not change if we only consider the secure sources (see Fig.~\ref{fig:LFCII}). Given the shape of the number density distribution, the \CC bins are also well fit by a power-law with slope $-0.94\pm0.21$ with $\chi^2_{\rm red}=0.4$. For most binning choices we explored, the $\chi^2_{\rm red}$ for the power-law and the Schechter fit were within $10$ per cent of each other.
\begin{figure}
\centering
\includegraphics[trim=0cm 0cm 0cm 0cm, width=0.479\textwidth]{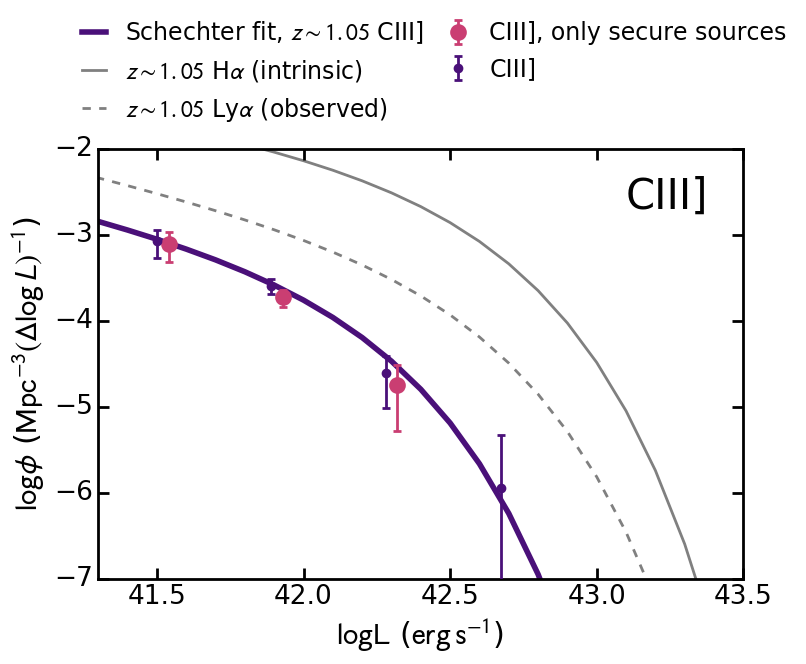}
\caption{Same as Fig.~\ref{fig:LFCII}, but for \C emitters. The \C LF at $z\sim1.05$ is well described by a Schechter function, consistent with a SF origin for the line emission. The \C LF is below Ly$\alpha$ and H$\alpha$ at all luminosities.}
\label{fig:LFCIII}
\end{figure}

\renewcommand{\arraystretch}{1.3}
\begin{table}
\begin{center}
\caption{Schechter LF parameters as resulting from fits. The faint end slope is fixed to a value of $-1.75$ for the Schechter fit. For \CC emitters we do not sample the bright end very well. \C densities are well described by a Schechter function. We were not able to get a converging Schechter fit for the \CIV data, as there is an infinitely large number of Schechter fits with $L^*$ higher than what we can probe that fit the data well, and that are thus indistinguishable from a power-law.}
\begin{tabular}{l c c c c c }
\hline\hline
Line & $z$ & $\alpha$ & $\log \phi^*$ & $\log L^*$  \\ 
              &     &          & (Mpc$^{-3}$)  & (erg\,s$^{-1}$)  \\ \hline          
\CC & $0.673-0.696$ & $-1.75$ & $-5.19^{+0.21}_{-0.40}$ & $42.79^{+0.58}_{-0.27}$\\
\C  & $1.039-1.066$ & $-1.75$ & $-3.60^{+0.12}_{-0.12}$ & $41.95^{+0.06}_{-0.06}$  \\ \hline
\end{tabular}
\label{tab:LF}
\end{center}
\end{table}
\renewcommand{\arraystretch}{1.1}

\subsection{\C luminosity function}\label{sec:LFCIII}

The \C LF is well fit by a Schechter function ($\chi^2_{\rm red}=0.4$, see Table~\ref{tab:LF} and Fig.~\ref{fig:LFCIII}). Given the large number densities and smaller Poissonian errors, the \C Schechter fit is the most secure out of all three Carbon emitters types, having the smaller errors on the LF parameters. The robustness of the bright sources above $L_{\rm CIII]}\sim10^{42.2}$ erg\,s$^{-1}$ is confirmed through spectroscopy or photometric redshifts. Note the high $z_{\rm spec}$ and $z_{\rm phot}$ completeness of \C emitters with 34 confirmed sources and the low overall expected fraction ($10-15$ per cent) of \C emitters compared to the entire emitter population selected with the CALYMHA NB survey. Because of this, including or excluding the uncertain sources through fractions does not significantly affect the results.

For completeness, we also fit a power-law function to the \C data which for most of the binning choices does not converge and in all other cases the $\chi^2_{\rm red}$ is a factor of a few worse than the Schechter fit.

\subsection{\CIV luminosity function}\label{sec:LFCIV}

A Schechter function with $\alpha$ fixed to $-1.75$ fits poorly the \CIV data at $z\sim1.5$. Due to the lack of a drop in number density at bright luminosities attempting to fit Schechter function resulted in a completely unconstrained characteristic $L^*$ and density $\phi^*$. Our minimisation did not converge and we were not able to find a $\chi^2$ minimum over the wide parameter space we probed (from $\log \phi^*$ of $-6$ to $2$ and $\log L*$ in the $40-46$ range). 

\begin{figure}
\centering
\includegraphics[trim=0cm 0cm 0cm 0cm, width=0.479\textwidth]{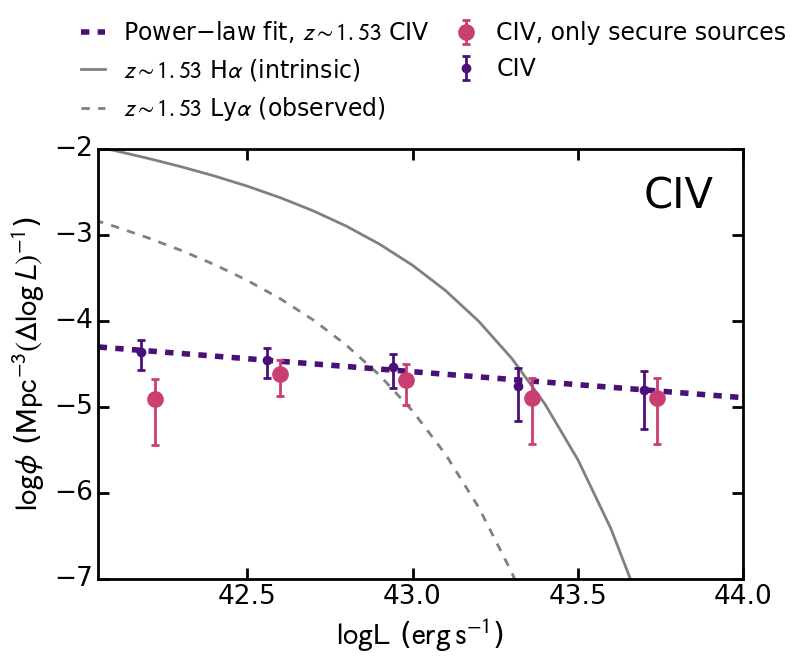}
\caption{\CIV emitter LF, fit with a power law. All labels similar to similar to Fig.~\ref{fig:LFCII}. A Schechter function could not be fit to the data. The results imply ubiquitous joint detections of Ly$\alpha$, H$\alpha$ and \CIV for bright sources.}
\label{fig:LFCIV}
\end{figure}

\renewcommand{\arraystretch}{1.3}
\begin{table}
\begin{center}
\caption{Power-law LF parameters, according to the fit described in equation \ref{eq:powerlaw}. \CC emitters are well by either a Schechter function or a power law. This is in line with the interpretation that \CC emitters are mainly powered by SF at lower fluxes and AGN the bright end. \CIV has a distinct power-law shape, characteristic of quasars.}
\begin{tabular}{l c c c c c}
\hline\hline
Line & $z$ & $\gamma$ & $\log L_{\rm int}$  \\ 
    &     &          & (erg\,s$^{-1}$) \\ \hline          
\CC & $0.673-0.696$ &  $-0.92\pm0.17$ & $34.15\pm6.96$  \\
\CIV & $1.513-1.546$ & $-0.30\pm 0.06$ &  $8.41\pm2.49$ \\ \hline 
\end{tabular}
\label{tab:LFpowerlaw}
\end{center}
\end{table}
\renewcommand{\arraystretch}{1.1}

\renewcommand{\arraystretch}{1.2}
\begin{table*}
\centering
\begin{threeparttable}
\caption{Cosmic average ratios of \CC, \C and \CIV to H$\alpha$ and Ly$\alpha$.  We compare to observed and intrinsic Ly$\alpha$ luminosity densities, corrected for escape fraction. The Ly$\alpha$ luminosity density was corrected for the power-law contribution expected at the bright end, as per \citet{Sobral2017}. We integrate the Schechter LFs fully and down to our observed limit to obtain the luminosity density $\rho$. In the case of the power-law fits, the densities depend on the integration limits, so we restrict the estimation to the range where we have directly measured the LF. Given the uncertainties in the LF fits and the estimation of the H$\alpha$ and Ly$\alpha$ LFs interpolated at our redshifts of interest, it is difficult to estimate the errors on the cosmic ratios. For the Schechter fits, we therefore report the cosmic ratio errors as obtain from departing the \CC and \C densities $\rho$ within its errors. For estimating errors, we assume that the H$\alpha$ and Ly$\alpha$ luminosity densities are known precisely. Therefore, errors on the ratios are underestimated.}
\begin{tabular}{l c c c c c c c}
\hline\hline
Line & $z$ & Fit type & $\log \rho$ &  $L$ range & \multicolumn{3}{c}{Cosmic line ratio}\\ \cmidrule{6-8}
              &     &          & (erg\,s$^{-1}$\,Mpc$^{-3}$)   & (erg\,s$^{-1}$) &  C/H$\alpha$ (observed) &  C/Ly$\alpha$ (observed) &   C/Ly$\alpha$ (intrinsic) \\ \hline          
\multirow{3}{*}{\CC} & \multirow{3}{*}{$0.673-0.696$} & Schechter & $38.16^{+0.62}_{-0.44}$ &  full &  $0.016^{+0.052}_{-0.011}$ & $0.09^{+0.28}_{-0.06}$ & $0.002^{+0.006}_{-0.001}$ \\
& & Schechter & $37.94^{+0.62}_{-0.44}$ & $10^{41.0}-\infty$ &  $0.014^{+0.045}_{-0.009}$ & $0.14^{+0.43}_{-0.09}$ & $0.002^{+0.005}_{-0.001}$ \\ 
& & Power-law & $38.18$ &  \phantom{000}$10^{41.0}-10^{43.0}$ & 0.025 & 0.24 & 0.003  \\ \hline
\multirow{2}{*}{\C}  & \multirow{2}{*}{$1.039-1.066$} & Schechter & $38.91^{+0.13}_{-0.13}$ &  full & $0.054^{+0.019}_{-0.014}$ & $0.24^{+0.09}_{-0.06}$ & $0.006^{+0.002}_{-0.002}$ \\ 
  & & Schechter & $38.22^{+0.13}_{-0.13}$ & $10^{41.5}-\infty$ & $0.022^{+0.008}_{-0.006}$ & $0.14^{+0.05}_{-0.04}$ & $0.003^{+0.001}_{-0.001}$ \\  \hline
\multirow{3}{*}{\CIV} & \multirow{3}{*}{$1.513-1.546$} & Schechter$^\dagger$ & $38.80$ & full & $0.025$ & $0.11$ & $0.003$  \\ 
& & Schechter$^\dagger$ & $38.79$ & \phantom{000}$10^{42.0}-\infty$ & $0.056$ & $0.46$ & $0.006$ \\ 
& & Power-law & $38.97$ & \phantom{000}$10^{42.0}-10^{43.5}$ & 0.086 & 0.71 & 0.010  \\ \hline
\end{tabular}
\begin{tablenotes}
\small
\item $^\dagger$ Given the very flat power-law fit to the \CIV LF, we decided, for the purposes of calculating cosmic densities, to also fit a Schechter fit. This avoids overestimating the luminosity density through the contribution of rare sources in a luminosity regime we are not directly probing.
\end{tablenotes}
\label{tab:cosmicratios}
\end{threeparttable}
\end{table*}
\renewcommand{\arraystretch}{1.1}
The LF is well described by a power-law with $\beta\sim0.3$ ($\chi^2\sim0.1$, see Fig.~\ref{fig:LFCIV}), similarly to the LF of quasars which are also power-law like (see also Section~\ref{sec:LFquasar}). Using the slope of the power-law fit as input for the faint end slope of the Schechter function, we get a fit of similar $\chi^2$ to the power law fit, however with an $L^*>10^{44}$ erg\,s$^{-1}$, which is beyond the range we can probe. Over the range of luminosities we measure, families of best-fit Schechter functions are indistinguishable from a power-law, while there is a single, well determined solution for a power-law. Hence, a power law is a simpler fit to the data. 

We note that not including unclassified sources through fractions slightly changes the values of the density bins. The most affected sources are those fainter in the BB. As discussed in Section~\ref{sec:select}, it is not surprising that fainter \CIV sources might not have photometric redshifts. At the bright end however, most sources are spectroscopically confirmed. The flat power law fit ($\gamma=-0.30\pm0.06$) might suggest the existence of \CIV sources beyond $L_{CIV}\sim10^{43-44}$ erg\,s$^{-1}$. However given the volume of our survey, the power-law LF indicates that at maximum 1 source per luminosity bin can be expected, which is in line with our non-detections beyond $10^{44}$ erg\,s$^{-1}$. We have also fitted a power law by excluding the lower luminosity bin which might be affected by incompleteness and found that the fit parameters are perfectly consistent with those from the fit with all bins. A NB survey of a larger volume would clarify the number densities of the brightest emitters.

\subsection{Comparison with H$\alpha$ and Ly$\alpha$}

We interpolate between the H$\alpha$ LFs at $z=0.2$ from \citet{2015MNRAS.453..242S} and $z\sim0.84,1.47,2.23$ from \citet{2013MNRAS.428.1128S} to the redshifts of our \CC, \C and \CIV emitters. We leave out the $z\sim0.4$ data point from  \citet{2013MNRAS.428.1128S}, since the volume was small, resulting in a larger $L^*$ uncertainty. \citet{2016A&A...591A.151G} present a sample of faint H$\alpha$ emitters at a very similar redshift to our \CC sample ($z\sim0.68$ versus $0.62$). The authors constrain the faint end slope of the H$\alpha$ LF at $z\sim0.62$ to $-1.41$ which is between the values at $z\sim0.2$ measured by \citet{2008ApJS..175..128S} and that at $0.8$ measured by \citet{2013MNRAS.428.1128S} and close to the value we derive through interpolation ($-1.46$ versus $-1.51$). Note that these LFs are corrected for intrinsic dust extinction of the H$\alpha$ line, as well as for all incompleteness.

For Ly$\alpha$, we interpolate to our reference redshifts, using the results at $z\sim0.3$ from \citet{2010ApJ...711..928C}, $0.9$ \citep{2012ApJ...749..106B}, $\sim2.2$ from \citet{2016ApJ...823...20K} and \citet{Sobral2017} and $3.1$ and $3.7$ from \citet{2008ApJS..176..301O}. We corrected the $z\sim0.9$ to fall on the expected evolution of $\phi^*$ and $L^*$ as shown in \citet{2016ApJ...823...20K}. We note that our Ly$\alpha$ LFs are observed quantities. The effect of the escape fraction of Ly$\alpha$ is further discussed in Section~\ref{sec:ratios}. Note that we do not show the Ly$\alpha$ power law component at high luminosities \citep{2016ApJ...823...20K,Sobral2017, Matthee2017}, but we do include it in the luminosity densities in the next section. The interpolated H$\alpha$ and Ly$\alpha$ LF parameters can be found in Table~\ref{tab:LFinterp}. 

\begin{figure*}
\centering
\includegraphics[trim=0cm 0cm 0cm 0cm, width=0.879\textwidth]{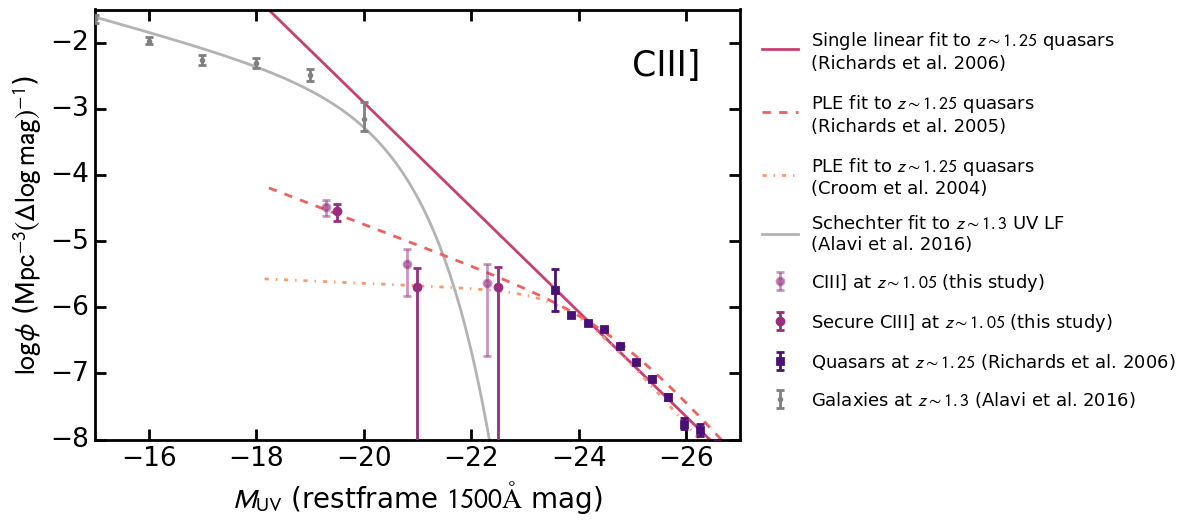}
\caption{Restframe $1500$\,{\AA} LF for \C emitters. We also show the bins obtained by using only secure sources, slightly shifted to the right for clarity. For comparison, we show the density of quasars as function of magnitude from \citet{2006AJ....131.2766R} and the galaxy UV LF from \citet{2016ApJ...832...56A}. About 2 per cent of faint galaxies with $M_{\rm UV}$ in the $-18$ to $-20$ range will be \C emitters, while UV bright galaxies and quasars with $M_{\rm UV}\gtrsim-22$ will have \C emission. Note that the characteristic UV luminosity $M^*$ at these redshifts is $\sim-20$, so these galaxies with \C emission are bright and rare.}
\label{fig:LFquasarCIII}
\end{figure*}

At fluxes corresponding to $L^*_{\rm CII]}$, the number densities of \CC emitters are consistent with the Schechter component of Ly$\alpha$ emitters at the same redshift, but significantly below those of H$\alpha$. If the power-law nature of the LF is valid for luminosities beyond those we can probe with our data, that would imply that bright \CC emitters could be just as numerous than H$\alpha$ and Ly$\alpha$, which also have strong power-law components. This however is quite surprising and physically hard to explain. It is thus likely that a Schechter function would provide a better fit at the brightest luminosities which we cannot probe with our current data. Assuming that H$\alpha$ and Ly$\alpha$ at this redshift are mostly produced in SF galaxies, the prevalence of bright \CC could indicate a different origin, that of AGN powering.

At $z\sim1$, our $L^*$ \C emitters have 20 per cent of the number density of Ly$\alpha$ and only about 3 per cent that of H$\alpha$ emitters. This could indicate that \C emission originates in SF galaxies such as those producing Ly$\alpha$ and H$\alpha$, but affected by typical line ratios expected from SF galaxies of a few per cent in relation to e.g. H$\alpha$.

At the brightest luminosities, given the very flat LF, \CIV number densities exceed those predicted by the Schechter component of the H$\alpha$ and Ly$\alpha$, further suggesting a quasar origin for the emission. At $z\gtrsim1$, the Ly$\alpha$ and H$\alpha$ LF have a power-law distribution beyond $L^*$, similar to the \CIV distribution \citep[e.g.][]{2016MNRAS.457.1739S, Matthee2017}. This is consistent with the typical joint detections of Ly$\alpha$, H$\alpha$ and \CIV tracing AGN at bright luminosities.

\subsection{Observed cosmic average line ratios to H$\alpha$ and Ly$\alpha$}\label{sec:ratios}

We integrate the LFs to obtain cosmic average ratios with respect to H$\alpha$ and Ly$\alpha$. These are useful for estimating the average relative emission line ratios of \CC, \C and \CIV emitters to H$\alpha$ and Ly$\alpha$ and to compare with theoretical predictions from AGN, SF models and with observations of individual sources. 

The Ly$\alpha$ LF has been shown to have a power-law bright end tail which for simplicity we did not included in our LFs \citep{2016ApJ...823...20K, Matthee2017, Sobral2017}. However, for the purpose of luminosity densities and cosmic ratios, this contribution is important. We therefore use the results at $z\sim2.2$ from \citet{Sobral2017}, to estimate as function of limiting integration luminosity, how much the power-law component contributes to the total luminosity density of Ly$\alpha$. We use these values to correct our Schechter $\rho_{Ly\alpha}$.

Ly$\alpha$ is scattered by neutral hydrogen and/or easily absorbed by dust and thus only a fraction escapes the galaxies. \citet{Sobral2017} computed the cosmic average Ly$\alpha$ escape fraction of $\sim5$ per cent at $z\sim2.2$ for the $3$ arcsec apertures we are using in this paper \citep[see also e.g.][]{2010Natur.464..562H}. This was achieved by comparing the ratios of the Ly$\alpha$ and H$\alpha$ luminosity densities versus the case B recombination value of 8.7. Since the escape fraction has been shown to evolve with redshift, we use the same method as \citet{Sobral2017} to estimate the Ly$\alpha$ escape fraction at our redshifts of interest using the interpolated LFs and find values of $1-2$ per cent. These are in line with the redshift dependent parametrisation of the Ly$\alpha$ escape fraction that \citet{2011ApJ...730....8H} derived using UV and emission line luminosity functions. We use our derived escape fractions to correct the observed luminosity densities of Ly$\alpha$ to intrinsic ones. We therefore also obtained line ratios between \CC, \C and \CIV to intrinsic values of Ly$\alpha$.

We list the luminosity densities in Table~\ref{tab:cosmicratios}, where we integrate the LFs fully and also within ranges probed directly by our data. 

Note that the cosmic ratio values are quite uncertain, because of the unknowns in deriving the LF parameters as well as the interpolation performed for H$\alpha$ and Ly$\alpha$ to obtain LFs at our redshifts of interest. For the Schechter fits, we derive errors as ranges in allowed by the error bars of the luminosity densities $\rho$. Note that we assume that the H$\alpha$ and Ly$\alpha$ densities are perfectly known, so the errors reported for the comic ratios are underestimated.

We find a typical ratio between \CC and H$\alpha$ of $\sim0.02$, while the observed \CC to Ly$\alpha$ ratio is higher at $\sim0.1$. This latter value is higher than the average for quasars which is $0.002$ \citep{2001AJ....122..549V}. \CC is therefore very weak in quasars compared to Ly$\alpha$, indicating our \CC are not quasars, but probably slightly less active AGN. 

The average \C line is weak compared to H$\alpha$ (ratio of $0.02-0.05$), in line with expectations for SF and well below AGN predictions from {\sc cloudy} (v 13.03) photo-ionisation modelling (Alegre et al. in prep). However, \C is non-negligible compared to observed Ly$\alpha$ with a ratio of $\sim0.1-0.2$. The observed cosmic average line ratio of \C to Ly$\alpha$ is consistent with the average for quasars \citep[0.16, ][]{2001AJ....122..549V} and a bit lower compared to results from $z\sim2-3$ and $z\sim6-7$ studies \citep{2010ApJ...719.1168E,2015MNRAS.454.1393S,2017MNRAS.464..469S}. See however our discussion in Section~\ref{sec:discussion}, where we show that once we take into account the Ly$\alpha$ escape fraction the values are consistent. Especially at the bright end where the quasar ratios are measured, \C might therefore be produced in AGN, however at the faint-end, another powering source, such as SF, would be necessary to maintain such a high cosmic ratio to Ly$\alpha$. 

\CIV at $z\sim1.5$ is weak compared to H$\alpha$ (ratios of $<0.1$). Our \CIV/H$\alpha$ ratios are much higher than those implied for SF from {\sc cloudy} modelling (with expected values of $0.003$, Alegre et al. in prep). Our average ratios to Ly$\alpha$ are relatively large, in the range of $0.1-1$, depending on the range for integration, but consistent with those measured for quasars \citep[0.25, ][]{2001AJ....122..549V}. This further supports a scenario where \CIV is mainly powered by AGN. If the high number densities of bright \CIV compared to Ly$\alpha$ are maintained at high redshift, large area surveys of \CIV could find suitable bright candidates for spectroscopic follow up.

\begin{figure*}
\centering
\includegraphics[trim=0cm 0cm 0cm 0cm, width=0.879\textwidth]{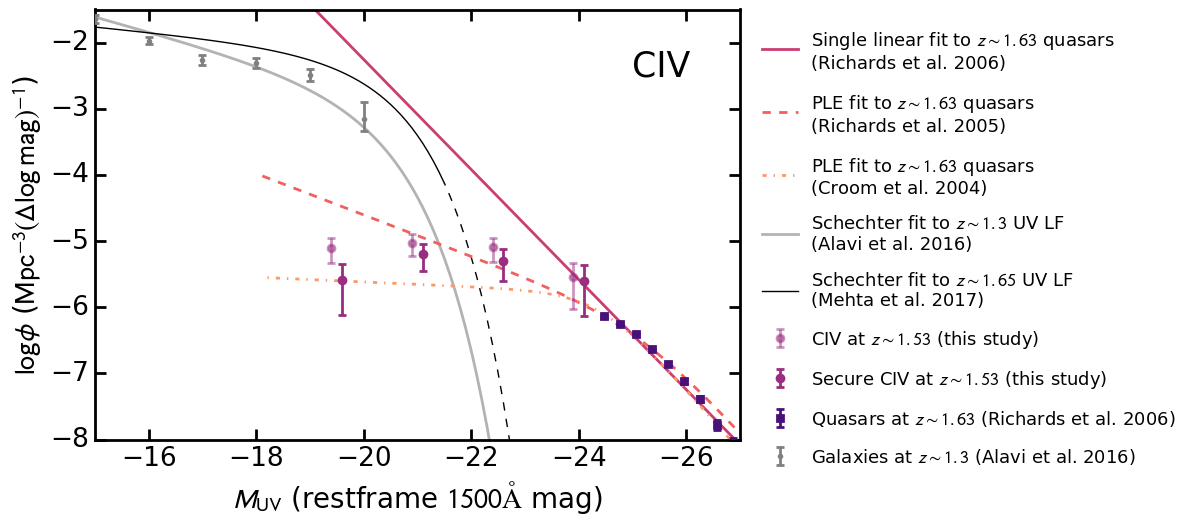}
\caption{Same as Fig.~\ref{fig:LFquasarCIII}, but for \CIV emitters. Note that not including the sources added through fractions results in incompleteness at faint magnitudes, where potential \CIV sources do not have photometric or spectroscopic redshifts. For the UV LF from \citet{2017ApJ...838...29M}, we plot the range directly probed by data in a solid line and the extrapolation of the Schechter fit in a dashed line. \CIV occurs in only $0.1-0.5$ per cent of fainter UV galaxies (galaxies at around $M^*$, i.e. $M_{\rm UV}\sim-20$), however bright sources will universally have \CIV emission with fluxes $>10^{-16.2}$ erg\,s$^{-1}$\,cm$^{-2}$.}
\label{fig:LFquasar}
\end{figure*}

\subsection{Comparison with the quasar and galaxy UV luminosity functions}\label{sec:LFquasar}

Motivated by the distinct quasar-like properties unveiled for our \CIV and possible AGN nature of bright \C emitters, we further investigate their nature by comparing their UV magnitudes to other galaxy/AGN populations.

We build a \C and \CIV restframe UV LF, using the method described in Section~\ref{sec:methods}. We use $U$ band (equivalent to about 1500\,{\AA} restframe wavelength) as all our emitters have a measured magnitude in this band. We apply our line flux completeness curve and restrict the LF to the range where the $U$ detections are $100$ per cent complete, i.e. apparent magnitudes brighter than $25$\footnote{This was chosen following \url{http://terapix.iap.fr/cplt/T0007/table_syn_T0007.html}}. The resulting \C LF can be seen in Fig.~\ref{fig:LFquasarCIII} and the \CIV LF in Fig.~\ref{fig:LFquasar}. 

For comparison, we use the binned quasar values from \citet{2006AJ....131.2766R} at $z\sim1.25$ and $z\sim1.63$ for \C and \CIV, respectively, transformed into the restframe $1500$\,{\AA} magnitude using the relation provided in the paper (eq. 3). We also compare to the single power-law fit from \citet{2006AJ....131.2766R} and the broken power law fits from \citet{2004MNRAS.349.1397C} and \citet{2005MNRAS.360..839R}, which assume a pure luminosity evolution (PLE). Over the range probed by the data from \citet{2006AJ....131.2766R}, all three models fit well, as their predictions differ mostly at the faint end. We also overplot the UV galaxy LF at $z\sim 1.3$ from \citet{2016ApJ...832...56A} and at $z\sim1.65$ from \citet{2017ApJ...838...29M}.

In the case of \C, including the sources added through fractions does not significantly affect the results. Note however that since we used observed $U$ band data for \C, this translates to a slightly redder restframe magnitude than the data we use for comparison. Our \C UV LF is slightly below the PLE fit to quasar densities \citep{2006AJ....131.2766R}, but probes a much fainter regime characteristic of SF galaxies. At magnitudes below $M_{\rm UV}\sim-21.5$, the fraction of SF galaxies that have \C above our EW threshold drops from 100 per cent at $M_{\rm UV} = -21.6$, down sharply to about 2 per cent at $-19.5$ mag. This means that UV sources which are intrinsically bright, with UV luminosities brighter than $M_{\rm UV}=-22$ will also be \C emitters. These results could further support the scenario emerging from the optical, X-ray and FIR data presented in \citet{PaperI} that \C is consistent with SF, but the brightest emitters are powered by AGN. 

When compared to the quasar luminosity densities inferred from the large, but shallow SDSS survey from \citet{2006AJ....131.2766R}, our \CIV data match well, but cover fainter restframe UV magnitudes. Our \CIV data points best match a broken power law PLE LF. Not including the sources with uncertain \CIV nature, which are added through fractions, affects the density values. This leads to a flattening at the faint magnitudes, further supporting our supposition in Section~\ref{sec:LF} that the sources without photometric redshifts are preferentially those which are faint in the continuum. Some of our fainter (UV luminosities $M_{\rm UV}$ from $-19$ to $-22$) \CIV emitters are in the galaxy regime probed by \citet{2016ApJ...832...56A}, but have much lower number densities (at least 2 orders of magnitude lower at magnitudes fainter than $M_{\rm UV}\sim-20$). The typical ratios to the UV galaxy densities at magnitudes fainter than $M_{\rm UV}\sim-21$ are of the order of $0.1-0.5$ per cent. This means that \CIV will only be produced in a very small fraction of faint galaxies at $z\sim1.5$. At faint magnitudes, \CIV might therefore be produced in SF but in very small fractions. This is in line with theory which suggests massive stars can contribute modestly to \CIV production \citep[e.g.][]{2016MNRAS.456.3354F}. The difference between the \CIV and Lyman break, UV selected LF further suggests \CIV is not a common occurrence in all galaxies and is not produced in large quantities in normal, SF galaxies at least at $z\sim1.53$. The bright magnitude distribution however, indicates that \CIV traces a very similar population to quasars.

\section{Discussion}\label{sec:discussion}

After unveiling the nature of \CC, \C and \CIV emitters from the point of view of individual multiwavelength properties in \citet{PaperI}, here we investigate their abundance by comparing to similar measurements for better studied emission lines, along with the galaxy and quasar UV luminosity functions.

\subsection{\CC emitters at $z\sim0.68$}

The \CC LF can be fit with either a Schechter-like LF or a power-law fit, which might suggest that \CC emitters may be a mix of a SF population and AGN activity (Fig.~\ref{fig:LFCII}). Especially at the bright end, a power-law shape of an LF has been shown to describe the data better in the case of mixed population of SF at fainter fluxes and AGN at bright fluxes as in the case of high-redshift Ly$\alpha$ and H$\alpha$ emitters \citep[e.g. $z\sim1-3$][]{2016ApJ...823...20K,2016MNRAS.457.1739S, Sobral2017, Matthee2017}. As we discussed in \citet{PaperI}, \CC emitters at fainter fluxes are most likely dominated by SF, given their UV and optical colours and morphologies, while the bright \CC emitters have Seyfert-like optical morphologies, with a disk and a bright nucleus and also have X-ray detections, which clearly indicate they host an AGN. Therefore, the mixed population nature of \CC emitters at $z\sim0.68$ is not only supported by individual properties, but is also revealed in the relative numbers of emitters found at each luminosity range. 

The number densities of \CC sources are much smaller than Ly$\alpha$ and H$\alpha$ (Fig.~\ref{fig:LFCII}). The cosmic average line ratio of \CC is only about $1-2$ per cent of H$\alpha$ and about $2$ per cent of the intrinsic Ly$\alpha$ (Table~\ref{tab:cosmicratios}, Fig.~\ref{fig:LFCII}). At higher redshifts a power law tail is indeed observed in the distribution of bright H$\alpha$ and Ly$\alpha$ emitters and can be associated with AGN. However, this effect has not yet been seen at lower redshifts with large surveys designed to capture the bright end well \citep[e.g. H$\alpha$ LF at $z\sim0.2$ and $0.8$,][]{2015MNRAS.453..242S, 2013MNRAS.428.1128S}. This might indicate that at these moderate redshifts, \CC can be significantly triggered in the shocks around more evolved, `maintenance-mode' galaxies which are accreting at smaller rates \mbox{\citep[e.g. such as active galaxies with radio jets][]{2000A&A...362..519D}}. This is also supported by the very high ratio of \CC to Ly$\alpha$, which is high when compared to averages for quasars. This again might suggest that \CC is not triggered in the quasar phase. We could interpret as a delayed effect of AGN activity: \CC might respond much at a phase where the nucleus is not actively accreting, but is in a maintenance mode. A large, shallow NB survey capturing the brightest, rarest \CC emitters and spectroscopic follow-up will allow significant progress. 

\subsection{\C emitters at $z\sim1.05$}

In \citet{PaperI} we showed that the vast majority of sources selected through their \C emission trace a similar population to H$\alpha$ selected SF galaxies with a range of properties (potentially including both low dust and dust-obscured sources). In this work, we find that a Schechter function fits the number densities of \C emitters well (Fig.~\ref{fig:LFCIII}). This suggests a SF nature for these sources and reinforces our results from \citet{PaperI}.

The LF is close to a scaled down version of the H$\alpha$ (by a factor of $\sim0.05$) and Ly$\alpha$ (by a factor of $\sim0.3$) LFs at $z\sim1.05$, which might indicate that \CC emission is produced in the same type of SF galaxies, but at a lower level (Fig.~\ref{fig:LFCIII}, Table~\ref{tab:LF}, Table~\ref{tab:LFinterp}), in line with photoionisation models \citep[e.g.][]{2016MNRAS.462.1757G}. \C line emission is only about $5$ per cent of H$\alpha$ and $\sim20$ per cent of Ly$\alpha$ (Table~\ref{tab:cosmicratios}). Assuming \C traces SF galaxies in a similar fashion to H$\alpha$ or Ly$\alpha$, this indicates that down to the same flux limit, a \C survey will be incomplete compared to these other two lines. To be able to obtain a similar number of emitters, a survey targeting \C would therefore have to be deeper by a factor of a few to be able to detect as many galaxies as an H$\alpha$ or Ly$\alpha$ survey. 

There is mild evidence at the bright end, that \C can be produced at similar levels at H$\alpha$ and Ly$\alpha$, which leads us to suggest that bright SF galaxies will universally have \C emission (Figs.~\ref{fig:LFCIII}, \ref{fig:LFquasarCIII}). This is in good agreement with the follow-up of some of the brightest high redshift UV selected galaxies, that have revealed \C detections \citep{2014MNRAS.445.3200S, 2015MNRAS.450.1846S}.

Our average intrinsic \C/Ly$\alpha$ cosmic line ratio is $0.6$ per cent (Table~\ref{tab:cosmicratios}). It is likely that because our measurements are cosmic averages, they will be lower than measurements of individual galaxies. In order to compare to the literature \citep{2010ApJ...719.1168E,2015MNRAS.454.1393S,2017MNRAS.464..469S}, we transform the observed \C to Ly$\alpha$ ratios into intrinsic values by estimating the escape fraction. We do this by using the relationship between Ly$\alpha$ EW (which the papers list) and its escape fraction as derived in \citet{Sobral2017}. We find that in these single galaxies at $z\sim2.2,6-7.0$ the intrinsic \C/Ly$\alpha$ ratio is in the $1-5$ per cent range, but these are the few detections, and thus they represent the highest values over the galaxy population, meaning the majority of other galaxies have lower ratios. Our value is therefore lower than literature samples, however the differences can be explained once we take into account the numerous non-detections of \C in Ly$\alpha$ detected galaxies \citep[e.g.][]{2017MNRAS.464..469S}. Note that many of the \C upper limits were in galaxies with Ly$\alpha$ just as strong as in those with \C detections. Once accounting for the lower \C to Ly$\alpha$ fractions of these galaxies undetected in \C, our results are in agreement with the literature. 

The ratio of \C to observed Ly$\alpha$ in our SF galaxies is similar to that of quasars. We have to keep in mind however, as discussed in the introduction, that Ly$\alpha$ escape fraction is expected to be very high in quasars, which are young, dust-free objects with a blue UV continuum \citep[e.g.][]{1988ApJ...327..570B, 2001ApJ...556...87H}. The intrinsic ratio of \C to Ly$\alpha$ in our galaxies is therefore probably smaller than in quasars. These perceived inconsistencies between cosmic ratios further reveal the stringent need for estimates of the Ly$\alpha$ escape fraction (as discussed in the introduction), as observed ratio to Ly$\alpha$ cannot be interpreted directly without knowledge of the host galaxy type, powering sources of Ly$\alpha$, amount of dust etc.
 
\subsection{\CIV emitters at $z\sim1.53$}

The power-law shape of the LF, similar to quasars (Fig.~\ref{fig:LFCIV}), is consistent with the individual properties of the galaxies in the sample with point-like optical morphologies and high X-ray luminosities \citep{PaperI}. The power-law fit to the \CIV number densities has a slope which is flatter than the quasar UV LF function at bright UV luminosities ($M_{\rm UV}\lesssim-24$), which has a very steep slope of $\sim-3$ \citep{2004MNRAS.349.1397C, 2005MNRAS.360..839R, 2006AJ....131.2766R}. However, it has been shown that the quasar LF is better described by a broken power-law with a shallower faint-end. For example, \citep{2005MNRAS.360..839R} found a faint-end slope of $\sim-1.8$, while \citep{2004MNRAS.349.1397C} find $\sim-1.1$. These two values bracket very well our power-law \CIV slope ($-0.35$ which translates to $-1.35$ in the definition in these papers).

Our LFs indicate that, at bright fluxes, \CIV emitters are more preponderant than Ly$\alpha$ and H$\alpha$. For the fainter luminosities, \CIV is weak compared to H$\alpha$ and produced in only a fraction of UV selected SF galaxies, however a high fraction of \CIV emitters are located in the quasar range of UV luminosities (Figs.~\ref{fig:LFCIV}, \ref{fig:LFquasar}). Therefore, \CIV will be produced in a vast majority of bright quasars, but conversely \CIV is unlikely in normal SF galaxies. 

The cosmic average ratio of \CIV to the observed Ly$\alpha$ fluxes at $z\sim1.5$ is just below 1 (Table~\ref{tab:cosmicratios}, with values close to those measured for quasars. Given that all the results point towards a scenario where \CIV host are quasars, then not only the cosmic average ratio to observed Ly$\alpha$ will be similar to quasars, but also the intrinsic ones. 

Therefore, from the point of view of properties and distribution, our data lends greater support to an actively accreting, young AGN powering source for \CIV. As theoretically predicted \citep{2006agna.book.....O}, \CIV is most probably triggered in the broad line region of young AGN. At least some of the fainter \CIV emitters are located in the `galaxy' regime (Fig.~\ref{fig:LFquasar}), which in line with more recent work, could be excited in strong ionising field in young galaxies \citep{2016MNRAS.456.3354F}.

\subsection{Prospects for observations at higher redshift}

\C and \CIV have been suggested as a possible alternative for spectroscopically confirming high-redshift galaxies. Our results, however, indicate that \C and \CIV are intrinsically very weak compared to Ly$\alpha$. In the epoch of reionisation, \C and \CIV could be an avenue to pursue only in the case of extremely low Ly$\alpha$ fractions. For example, even for a $1-2$ per cent Ly$\alpha$ escape fraction as assumed in this paper, \C still has a cosmic ratio of only $0.2$ compared to Ly$\alpha$. A rough calculation indicates that only in the case of escape fractions $\lesssim 1$ per cent, would \C and \CIV become more efficient than Ly$\alpha$. However, in spite of the low effective Ly$\alpha$ escape fractions expected at high redshift, a source would still need to be intrinsically bright to have strong \C emission such that it can be spectroscopically followed up. 

Another point to consider is the population that \C and \CIV trace. A number of studies proposed that \C and \CIV can be used to trace low mass, young, metal-free galaxies at high redshifts \citep[e.g.][]{2014MNRAS.445.3200S, 2015MNRAS.450.1846S, 2015MNRAS.454.1393S, 2017ApJ...836L..14M, 2017ApJ...839...17S}. However, our results indicate that while \C emission indeed traces a general population of SF galaxies, \CIV at $z\sim1.5$ seems to be almost universally produced in quasars, not SF galaxies. We speculate that the galaxies detect in \C or \CIV at high redshift might be the ones with lower Ly$\alpha$ escape fraction (e.g. the ones with relatively low Ly$\alpha$ EWs).

It is therefore crucial to perform a blind survey similar to the one in this paper, but at high redshift to assess the redshift evolution of the distribution and nature of \CC, \C and \CIV.

\section{Conclusions}\label{sec:conclusion}

The brightest restframe UV lines after Ly$\alpha$, \C and \CIV have been proposed as a promising way to spectroscopically confirm and study high redshift galaxies in the epoch of reionisation, where Ly$\alpha$ is expected to become harder to observe. However, to date, no comprehensive survey has been performed to identify the nature and abundance of hard ionisation lines such as \CC, \C and \CIV. Using the uniformly selected samples from \citet{PaperI}, we estimate the statistical properties of \CC, \C and \CIV emitters for the first time. By studying lower redshift emitters, \CC at $z\sim0.68$, \C at $z\sim1.05$ and \CIV at $z\sim1.53$, we can obtain accurate photometric redshifts to distinquish from Ly$\alpha$ emitters and study their properties in greater details than possible at high redshift. Our results are as follows:
\begin{itemize}
\item The \CC LF is well by either a Schechter or a power law (Fig.~\ref{fig:LFCII}), consistent with a mixed population of SF at fainter fluxes and AGN at brighter fluxes in line with properties of individual sources from \citet{PaperI}. The \CC line emission is on average weaker than H$\alpha$ and Ly$\alpha$, with a cosmic average line ratio of $\sim0.015$ and $0.1$, respectively.
\item The \C LF is well described by a Schechter function with parameters  $\log \phi^*=-3.60\pm0.12$ and $\log L^* = 41.95\pm0.06$, when the faint end slope is fixed to $\alpha=-1.75$ (Fig.~\ref{fig:LFCIII}). While tracing a similar population of SF galaxies, \C emitters at $z\sim1.05$ are relatively rare compared to typical Ly$\alpha$ and H$\alpha$ emitting sources. While, the average line ratio of \C to H$\alpha$ is $0.05$, the ratio to Ly$\alpha$ is about $0.25$. Our results indicate that all bright UV selected galaxies as well as bright quasars will have \C emission ($M_{\rm UV}<-22$, Fig.~\ref{fig:LFquasarCIII}).
\item By contrast, the \CIV LF is best described by a single, quasar-like power-law (Fig.~\ref{fig:LFCIV}). As revealed by our data, \CIV is produced in only a small fraction of UV selected, H$\alpha$ and Ly$\alpha$ SF galaxies, but is pervasive in quasars  (Fig.~\ref{fig:LFCIV}, \ref{fig:LFquasar}). As expected from simulations, the cosmic average ratio of \CIV to Ly$\alpha$ is relatively large ($0.1-1$), with the highest ratios for the most luminous sources. 
\end{itemize}

To conclude, a consistent picture about the nature and powering source of \CC, \C and \CIV emission is emerging from the individual properties unveiled in \citet{PaperI} and the statistical behaviour studied here. As shown in \citet{PaperI}, \CC emitters at $z\sim0.68$ have morphologies and colours consistent with SF at lower luminosities, while bright sources X-ray detections reveal relatively active central black holes. The statistical properties of the sample, including the shape of the LF and the ratios to lines such as H$\alpha$ and Ly$\alpha$, confirm the scenario of a mixed population. At $z\sim1.05$, \C emitters have optical morphologies and colours characteristic of either isolated or interacting SF galaxies \citet{PaperI}. In many cases, the peak of the line emission is offset from the main UV region, which can explain the large EWs these emitters have. Interactions could for example be responsible for triggering \C emission away from the main stellar disk. In this work, we find that indeed \C emitters have a luminosity distribution consistent with a SF origin and they probably trace a similar population with H$\alpha$ emission. However, \C has a relatively low ratio to H$\alpha$. Through their relatively flat, power-law luminosity density distributions, point-like optical morphologies, blue colours, steep UV slopes and strong X-ray detections indicative of high black hole accretion, \CIV emitters at $z\sim1.53$ in our sample are quasars almost in entirety.

Our statistical study indicates that, while \CC, \C and \CIV are indeed not uncommon occurrences at $z\sim0.7-1.5$, they are much rarer than Ly$\alpha$ or H$\alpha$ emitters for a fixed flux limit. \C and \CIV lines might prove to be a suitable avenue for spectroscopic confirmation of the highest redshift galaxies only if Ly$\alpha$ is heavily attenuated and less than 1 per cent escapes within the slit aperture. Other concerns are raised regarding the population traced by \C and \CIV. While a \C selected sample might trace a part of the SF population, \CIV will most likely be biased towards quasars. Unless the properties of \C emitters evolve strongly with redshift, our prospects for \C and \CIV detections of typical SF galaxies at the highest redshifts are less optimistic compared to the small number statistics studies performed through lensed sources at $z\sim7$ by \citet{2015MNRAS.450.1846S} and simulations by \citet{2016MNRAS.461.3563S}. 

\section*{Acknowledgements}

We would like to thank the anonymous referee for her/his valuable input that helped improve the clarity and interpretation of our results. DS acknowledges financial support from the Netherlands Organisation for Scientific research (NWO), through a Veni fellowship. IO acknowledges support from the European Research Council in the form of the Advanced Investigator Programme, 321302, {\sc cosmicism}. CALYMHA data is based on observations made with the Isaac Newton Telescope (proposals 13AN002, I14AN002, 088-INT7/14A, I14BN006, 118-INT13/14B, I15AN008) operated on the island of La Palma by the Isaac Newton Group in the Spanish Observatorio del Roque de los Muchachos of the Instituto de Astrof{\'i}sica de Canarias. Also based on data products from observations made with ESO Telescopes at the La Silla Paranal Observatory under ESO programme IDs 098.A-0819 and 179.A-2005. We extensively used the cosmology calculator presented in \citet{2006PASP..118.1711W}. We would like to thank the authors of NumPy \citep{numpy}, SciPy \citep{scipy}, Matplotlib \citep{matplotlib} and AstroPy \citep{astropy} for making these packages publicly available. This research has made use of the NASA/IPAC Extragalactic Database (NED) which is operated by the Jet Propulsion Laboratory, California Institute of Technology, under contract with the National Aeronautics and Space Administration. This research has made use of NASA's Astrophysics Data System. This research has made use of the VizieR catalogue access tool, CDS, Strasbourg, France. The original description of the VizieR service was published in \citet{2000A&AS..143...23O}. This research has made use of ``Aladin sky atlas" developed at CDS, Strasbourg Observatory, France \citep{2000A&AS..143...33B,2014ASPC..485..277B}. 

\bibliographystyle{mn2e.bst}

\bibliography{CarbonEmittersII}

\appendix
\section{Interpolated H$\alpha$ and Ly$\alpha$ values}

We use data from the literature at redshifts in the $z\sim 0.2 - 3.1$ range to derive H$\alpha$ 
\citep{2013MNRAS.428.1128S, 2015MNRAS.453..242S} and and observed Ly$\alpha$ \citep{2010ApJ...711..928C, 2012ApJ...749..106B, 2016ApJ...823...20K, Sobral2017, 2008ApJS..176..301O}. LFs at our redshifts of interest for \CC, \C and \CIV. Note that the values are for reference only and should not be considered as actual LF parameters at the interpolated redshifts.

\renewcommand{\arraystretch}{1.3}
\begin{table}
\begin{center}
\caption{The interpolated H$\alpha$ and observed Ly$\alpha$ Schechter LF parameters at the redshifts of the \CC, \C and \CIV emitters. We used the results from \citet{2013MNRAS.428.1128S} and \citet{2015MNRAS.453..242S} for H$\alpha$ and a range of papers for Ly$\alpha$ \citep{2010ApJ...711..928C, 2012ApJ...749..106B, 2016ApJ...823...20K, Sobral2017, 2008ApJS..176..301O}.}
\begin{tabular}{l c c c c }
\hline\hline
 $z$ & $\alpha$ & $\log \phi^*$ & $\log L^*$ \\ 
 &          & (Mpc$^{-3}$)  & (erg\,s$^{-1}$)  \\ \hline          
\multicolumn{4}{c}{Interpolated H$\alpha$} \\
$0.68$ & $-1.54$ & $-2.56$ & $42.22$ \\
$1.05$ & $-1.60$ & $-2.52$ & $42.35$   \\
 $1.53$ & $-1.60$ & $-2.62$ & $42.68$  \\\hline
\multicolumn{4}{c}{Interpolated Ly$\alpha$} \\
$0.68$ & $-1.80$ & $-3.60$ & $42.11$  \\
 $1.05$ & $-1.79$ & $-3.45$ & $42.30$ \\
 $1.53$ & $-1.78$ & $-3.31$ & $42.42$  \\
\hline
\end{tabular}
\label{tab:LFinterp}
\end{center}
\end{table}
\renewcommand{\arraystretch}{1.1}

\section{Number densities and number of sources for the LF fits}

\renewcommand{\arraystretch}{1.3}
\begin{table}
\begin{center}
\caption{The number densities in each bin for the \CC, \C and \CIV LFs, as well as the number of sources going into each bin, corrected for incompleteness and including fractions added for the sources with uncertain nature. The data points can be also visualised in Figs.~\ref{fig:LFCII},~\ref{fig:LFCIII} and ~\ref{fig:LFCIV}.}
\begin{tabular}{l c c}
\hline\hline
 $\log L$ & $\log \phi$ & $N_{\rm sources}$ \\ 
\hline
\multicolumn{3}{c}{\CC} \\
41.21	&	$-3.61^{0.12}_{-0.18}$ &	9	\\
41.60	&	$-4.06^{0.15}_{-0.23}$ &	6	\\
41.99	&	$-4.22^{+0.17}_{-0.30}$ &	4	\\
42.38	&	$-4.73^{+0.28}_{-1.13}$ &	1	\\
42.77	&	$-5.17^{+0.42}_{-\inf}$ &	1	\\
\hline
\multicolumn{3}{c}{\C} \\
41.52	&	$-3.08^{+0.14}_{-0.20}$ &	8	\\
41.91	&	$-3.59^{+0.08}_{-0.09}$ &	30	\\
42.30	&	$-4.61^{+0.20}_{-0.40}$ &	3	\\
42.69	&	$-5.94^{+0.62}_{-\inf}$ &	1	\\
\hline
\multicolumn{3}{c}{\CIV} \\
42.20	&	$-4.37^{+0.14}_{-0.20}$ &	7	\\
42.58	&	$-4.46^{+0.14}_{-0.21}$ &	7	\\
42.96	&	$-4.54^{+0.15}_{-0.24}$ &	6	\\
43.34	&	$-4.76^{+0.21}_{-0.41}$ &	4	\\
43.72	&	$-4.81^{+0.22}_{-0.45}$ &	3	\\
\hline
\end{tabular}
\label{tab:binsL}
\end{center}
\end{table}
\renewcommand{\arraystretch}{1.1}

\renewcommand{\arraystretch}{1.3}
\begin{table}
\begin{center}
\caption{The number densities in each bin for the \C and \CIV UV magnitude LFs. We also list the number of sources in each bin, which we correct for incompleteness and source fractions added for the sources with uncertain nature. The data points can be also visualised in Figs.~\ref{fig:LFquasarCIII} and~\ref{fig:LFquasar}.}
\begin{tabular}{l c c}
\hline\hline
 $M_{\rm UV}$ & $\log \phi$ & $N_{\rm sources}$ \\ 
\hline
\multicolumn{3}{c}{\C} \\%
-22.4	&	$-5.63^{+0.28}_{-1.10}$ &	1	\\
-20.9	&	$-5.35^{+0.22}_{-0.49}$ &	2	\\
-19.4	&	$-4.48^{+0.10}_{-0.13}$ &	17	\\
-17.9	&	$-4.25^{+0.08}_{-0.10}$ &	28	\\
\hline
\multicolumn{3}{c}{\CIV} \\
-24.0	&	$-5.55^{+0.22}_{-0.48}$	 &	2	\\
-22.5	&	$-5.10^{+0.15}_{-0.22}$	 &	6	\\
-21.0	&	$-5.03^{+0.14}_{-0.20}$	 &	7	\\
-19.5	&	$-5.11^{+0.15}_{-0.23}$	 &	6	\\
-18.0	&	$-5.23^{+0.17}_{-0.29}$ &	5	\\
\hline
\end{tabular}
\label{tab:binsM}
\end{center}
\end{table}
\renewcommand{\arraystretch}{1.1}

\end{document}